\documentclass[twocolumn,floatfix,a4paper]{revtex4-1}

\usepackage{epsf}
\usepackage{epsfig}
\usepackage{color}
\usepackage{psfrag}
\usepackage{verbatim}
\usepackage{amsmath}
\usepackage{amsfonts}
\usepackage{appendix}
\usepackage{graphicx}
\usepackage{dsfont}

\newcommand{\rmd}{{\rm d}}

\begin{document}

\allowdisplaybreaks


\title{The dissipative phase transition in a pair of coupled noisy two-level systems}

\author{Julius Bonart}
\affiliation{Laboratoire de Physique Th\'eorique et Hautes Energies,
Universit\'e Pierre et Marie Curie -- Paris VI, 
4 Place Jussieu, 75252 Paris Cedex 05, France
}

\begin{abstract}
We study the renormalization group (RG) equations of a pair of spin-boson systems coupled in the $z$-direction with each other. Each spin is coupled to a different bath of harmonic oscillators. We introduce a systematic adiabatic RG, which generalizes the first-order adiabatic renormalization previously used for the single spin-boson model, and we obtain the flow equations for the tunneling constant, the dissipation strength and the inter-spin coupling up to third order in the tunneling. If one of the two spins is treated as a constant magnetization the other spin is described by a biased spin-boson Hamiltonian. In this case the RG equations we find coincide with the ones obtained \emph{via} a mapping to a long-range Ising chain. If the whole \emph{Ohmic} two-spin system is considered the Kosterlitz-Thouless phase transition is replaced by a second-order phase transition. In the case of a \emph{sub-Ohmic} bath our approach predicts that the two-spin system is always localized. 
\end{abstract}

\maketitle


%


\section{Introduction}

The influence of dissipation on quantum systems can be profound. Fundamental principles in quantum mechanics such as Heisenberg's uncertainty relation, as well as basic phenomena such as quantum 
tunneling often exhibit a surprising interplay with the effects of external thermal or quantum noise~\cite{CalLegg81,GrabertWeiss84,Dorsey85,Zwerger85,SchrammGrabert86,Grscin88,Hanggi05}. One of the most important models for quantum dissipation is the so called 
``spin-boson model'' which describes a single spin coupled to a bosonic quantum bath~\cite{CalLegg87}. Its applications are numerous. Originally devised to model a particle in a double-well 
potential~\cite{Br02}, the spin-boson model widely occurs in the field of quantum optics and quantum computation where it is used to describe a noisy two-level system or qubit 
(see, e.g.~\cite{Brandes02, Martinis03, Bellomo07}). 
In its standard form, the spin-boson Hamiltonian for a noisy spin $\vec s = \frac{1}{2}(\sigma^x,\sigma^y,\sigma^z)$ with bias $\epsilon$ and tunneling rate $h$ reads 
\begin{equation}\label{eq:spinboson}
  \mathcal H_{\rm SB} = \frac{\epsilon}{2} \sigma^z + \frac{h}{2}\sigma^x + \sum_i \lambda_i(a^\dagger_i+a_i)\sigma^z + \sum_i\omega_i a^\dagger_i a_i \; ,
\end{equation}
where $a_i^\dagger,a_i$ are the ladder operators of the quantum bath, which is entirely characterized by its spectral density
\begin{equation}\label{eq:spectral}
  S(\omega) = \sum_i \delta(\omega_i-\omega)\lambda_i^2 = \frac{\alpha \omega^s}{2} \Theta(\omega-\omega_c)\; ,
\end{equation}
where $\alpha$ is the dissipation strength and $\omega_c$ a high-frequency cutoff that we choose to impose in a hard way. The so-called Ohmic case corresponds to $s =1$ while a spectral density 
with $s<1$ ($s>1$) is called sub-Ohmic (super-Ohmic, respectively). 
Despite its simple form, the model defined in~(\ref{eq:spinboson}) is not exactly solvable, although tremendous progress has been recently made for the single 
bath-mode case~\cite{Braak11}. 
If coupled to an Ohmic bath, $S(\omega) = \alpha\omega$ for $\omega$ small, it is well-known that at zero temperature the system (\ref{eq:spinboson}) has a quantum phase transition at 
$\alpha_c =1$ (to lowest order in $h/\omega_c$)~\cite{CalLegg87} which separates a localized regime (with zero tunneling probability) from the delocalized one (where tunneling occurs). There is now a 
consensus that the super-Ohmic bath does not induce any phase transition (at zero temperature) while the sub-Ohmic spin-boson model has a second-order 
transition between a localized and a delocalized phase separated by an $s$- and $h/\omega_c$-dependent critical $\alpha_c(s,h/\omega_c)$~\cite{Vojta05,CalLegg87}.

Systems consisting of more than one noisy two-level system have attracted much interest in recent years. 
At a macroscopic level, 
the ferromagnetic dissipative quantum Ising chain in which $N_s$ spin-boson units are placed on a one 
dimensional lattice and coupled  \emph{via} a ferromagnetic nearest-neighbour 
interaction has been numerically studied in~\cite{Troyer05,Cugl05}. The same system with an additional strong disorder has also been investigated in the past~\cite{Cugl05,Schehr08,Hoyos08,Hoyos12}. 
These studies concluded that the dissipative quantum spin chain does not lie in 
the $2d$ classical Ising universality class.
The same one-dimensional system with exchanges drawn from a probability distribution 
has been analyzed, also numerically, in~\cite{Cugl05}. 
However, systems with  a 
finite number of coupled noisy units are also interesting not only because they 
can be interpreted as a non-trivial part of the dissipative Ising chain but, 
more importantly, because in their minimal expression they are the simplest logical element. 
Indeed,  two qubits can form, for instance, a quantum exclusive OR gate. The interest in understanding the 
dynamics of such two coupled noisy spins is therefore huge and a large number of papers were devoted to the 
analysis of different aspects of them. Just to mention a few, decoherence and entanglement of \emph{two} coupled 
qubits have been studied in~\cite{Dube98,Hanggi01,Bellomo07}. Two qubits coupled to the \emph{same} bosonic environment have been considered in~\cite{LeHur10,CutcheonFisher10,Tornow08}. The dynamics of two-spin system in imperfect crystals have been investigated, e.g., in~\cite{Nalbach02}.

%

The critical behaviour of the Ohmic spin-boson model can be tackled by a variety of methods. 
The non-interacting blip approximation (NIBA) captures well the behaviour of the unbiased Ohmic dissipative two-level system (at least for moderate $\alpha$ and not too large times) but it fails in presence 
of a finite bias or a sub-Ohmic spectral density~\cite{CalLegg87}. Hence, it is probably not suited for the 
study of coupled spin-boson systems 
when the inter-spin coupling behaves \emph{de facto} as a finite bias. 
In the renormalization group (RG) approach the idea is that the renormalization of the tunneling characterizes the phase. 
More precisely, if $h$ renormalizes to zero one concludes that the spin is localized while
if $h$ remains finite it is delocalized. 
The spin-boson Hamiltonian (\ref{eq:spinboson}) can be mapped onto a long-range classical Ising chain 
with external field whose RG equations 
are known to lowest order~\cite{CalLegg87,Kosterlitz76}. By carrying over these results to the spin-boson model, 
one finds the Kosterlitz-Thouless equations~\cite{Vojta03} for $\alpha$ and $h$:
\begin{eqnarray}
\label{eq:RG1}
  \partial_\ell (\alpha\omega_c^{s-1}) &=& (1-s)\alpha\omega_c^{s-1} - \alpha\omega_c^{s-1} (h/\omega_c)^2 \label{eq:RG11a}\\
  \partial_\ell (h/\omega_c) &=& \left[1-\alpha\omega_c^{s-1}\right](h/\omega_c)\;. \label{eq:RG11}
\end{eqnarray}
In~\cite{Chakravarty82} an additional equation for the bias is reported which takes the form (we use our notations and corrected a typo) 
\begin{equation}
  \partial_\ell (\epsilon/\omega_c) = \left[1 - \frac{1}{2}(\alpha\omega_c^{s-1})(h/\omega_c)^2\right](\epsilon/\omega_c) \; . 
\end{equation}
Our aim is to go beyond the single spin-boson model and to study the properties of two coupled units.
To state the problem more precisely, let us assume that the two spins are coupled through 
a $z$-interaction constant $J$, such that their Hamiltonian reads
\begin{eqnarray}
  \mathcal H &=& \frac{J}{4}\sigma^z_1 \sigma^z_{2} 
  +
  \frac{h}{2}\sum_{n=1}^2 \sigma_n^x 
  \nonumber\\
&& 
+ \sum_{n=1}^2 \sum_{i_n=1}^{N_n} \lambda_{i_n}\left(a^\dagger_{i_n} + a_{i_n}\right) \sigma_n^z  
+
   \mathcal H_B
\label{eq:ham2s}
\end{eqnarray}
with 
\begin{equation}
\mathcal H_B = \sum_{n=1}^2 \sum_{i_n=1}^{N_n} \omega_{i_n}a^\dagger_{i_n}a_{i_n}
\end{equation}
and $N_n$ the total number of oscillators coupled to the $n$-th spin. As the baths are 
independent, the creation and annihilation operators with different $n$ commute with each other. Note that in the case where the two spins are coupled to a common bath, correlations between the spins are directly induced \emph{via} the bath~\cite{LeHur10}. The following analysis is heavily changed when correlations between the different baths exist and the results in this article thus apply only to the case where each spin is coupled to its own environment.

Now, if we treat one spin -- say the second one -- in a mean-field like way, the first spin feels the magnetization $m = \langle\sigma_2^z\rangle/2$ of the 
second spin \emph{via} $Jm\sigma_1^z/2$. Thus, within the mean-field approximation, this leads to a single-spin model with 
a finite bias $\epsilon = Jm$. The full (non mean-field) double spin-boson system could in principle be also mapped onto a classical model and studied from this point of view; however, we are not aware of any such analysis in the 
literature.

In this paper we show how a RG scheme can be constructed for the full noisy two-qubit system without using any mapping to a classical system. Our approach is inspired by the adiabatic renormalization scheme which aims at successively 
integrating out the high-energy bath modes. The first order adiabatic renormalization has already been applied to the unbiased single spin-boson 
Hamiltonian with success~\cite{CalLegg87} and the result is Eq.~(\ref{eq:RG11}) [the second-order equation (\ref{eq:RG11a}) were not derived in 
this way but with the mapping to a classical $2d$ system]. 
However, a systematic higher order analysis for the coupled two-spin system has not been performed yet. The objective of this paper is to close this gap 
and to demonstrate the power of the systematic adiabatic RG scheme. In particular, we will find the critical behaviour of the two-spin system and discuss its implications for the dissipative quantum Ising chain.%

This paper is organized as follows. In the first section we present the adiabatic renormalization scheme and we determine the RG equations 
for $\epsilon$, $h$ and $\alpha$ for a single noisy spin up to third order in $\tilde h\equiv h/\omega_c$. 
In the second 
section we consider two coupled spin-boson systems and we determine the RG equation for the inter-spin coupling $J$. By performing a small- and a large-$J$ analysis we derive the critical behaviour of the full coupled system. We show that both the nature of the Ohmic and the sub-Ohmic phase transition is different from the standard phase transition of the unbiased single-spin boson system.

\section{The adiabatic renormalization scheme}

The adiabatic renormalization scheme is, to our knowledge, the first method that has been successfully applied to the dissipative 
two-state system~\cite{Chakravarty82,CalLegg87,We08}. The idea is to exploit the separation of energy scales which occurs when the bath cutoff 
$\omega_c$ is much larger than both $h$ and $J$. In this case some bath modes are so highly energetic that they adapt 
instantly to the spin's state. 

To be more specific we consider a Hamiltonian 
\begin{eqnarray}\label{eq:ham21}
  \mathcal H &=& \mathcal H_S[\{\sigma^z_n\}] + 
\frac{h}{2}\sum_n \sigma_n^x + \sum_{n,i_n} \lambda_{i_n}\left(a^\dagger_{i_n} + a_{i_n}\right)\sigma_n^z \nonumber\\
&&+ \ \mathcal H_B \; .
\end{eqnarray}
$H_S[\{\sigma^z_n\}]$ is a generic interaction term between the $z$-components of the spins and has 
an energy scale of the order of $J$ that, by assumption, satisfies $J \ll \omega_c$. 
We take $\lambda_{i_n} \sim 1/\sqrt{N_n}$ thus ensuring  
an homogeneous scaling of the bath strength. All baths are characterized by the same spectral density that we take  
to be given by Eq.~(\ref{eq:spectral}). 

We now shift the high energy bath modes (more precisely, those with $\omega_c e^{-\rmd \ell}<\omega<\omega_c$ and $\rmd \ell$ a small parameter) 
to their equilibrium position conditioned on the spin state $\sigma_n^z = \pm 1$. It is straightforward to calculate the required unitary 
transformation~\cite{CalLegg87}:
\begin{equation}\label{eq:U}
  U = \exp \left[ \sum_{n,j_n,>}\frac{\lambda_{j_n}}{\omega_{j_n}} \left(a^\dagger_{j_n} - a_{j_n}\right) \sigma_n^z \right] \; ,
\end{equation}
where the subscript $>$ indicates that the sum runs over all high frequency bath modes only. 
It is known that the NIBA~\cite{CalLegg87} corresponds to a Born-Oppenheimer like approximation which is applied after the transformation~\ref{eq:U} with the sum now running over \emph{all} bath modes~\cite{Aslangul86}. However, it is also known that the corresponding zero-order wavefunction substantially differs from the true one for low frequencies. This has been pointed out in~\cite{Chin06} to be the main reason why NIBA fails for a sub-Ohmic bath. By transforming only the high-frequency bath modes according to~\ref{eq:U} we avoid this problem. 
The transformed Hamiltonian can then be written as 
\begin{equation}\label{eq:Hprime}
  \mathcal H' = U \mathcal H U^\dagger = \mathcal H_B^> + \mathcal V\;,
\end{equation}
with
\begin{eqnarray}
  \mathcal H_B^> &=& \sum_n \sum_{i_n,>}\omega_{i_n} a^\dagger_{i_n} a_{i_n}  \; ,\\
\mathcal V &=& \frac{h}{2} \sum_n \left( e^{v_n}\sigma_n^+ + e^{-v_n}\sigma_n^-\right)
+ \sum_n \sum_{i_n,<}\omega_{i_n} a^\dagger_{i_n} a_{i_n} 
\nonumber\\ 
&+& \sum_n \sum_{i_n,<} \lambda_{i_n}\left( a^\dagger_{i_n} + a_{i_n}\right) \sigma_n^z + H_S[\{\sigma^z_n\}]\;\label{eq:Vext}  ,
\end{eqnarray}
where the subscript $<$ constrains the sum to run only over bath modes with ``small'' frequencies 
$\omega<\omega_c e^{-\rmd \ell}$. We introduced the anti-Hermitian operator
\begin{equation}\label{eq:smallv}
v_n = \sum_{i_n,>} \frac{2\lambda_{i_n}}{\omega_{i_n}}\left( a^\dagger_{i_n} - a_{i_n} \right) \; ,
\end{equation}
that is nothing else than a sum of rescaled oscillator momentum operators (and which consequently shift the oscillators' positions). Up to this point our treatment is still exact. From now on we call $\mathcal H$ the transformed Hamiltonian in (\ref{eq:Hprime}). 

As the exact ground state of the Hamiltonian (\ref{eq:Hprime}) is unknown for finite $h$
we need to use other means to determine the behavior of the system.
By using degenerate perturbation theory we will find the approximate ground state
which, in return, will yield information on the quantum critical point.  
%
%
%
%
%
In the following we will describe  this \emph{adiabatic renormalization scheme} in detail.

%

First, we note that the eigenstates of $\mathcal H'$ have the entangled form $|\vec s,\vec n_>, \vec n_<\rangle$, with the 
$z$-spin states $s_n = \pm$, the fast oscillators with occupation numbers $\vec n_>$ and the slow oscillators with 
occupation numbers $\vec n_<$. The lowest energy state is the one with $\vec n_> = 0$.

Second, at each step we force only a fraction of oscillators -- namely the high energy ones -- to be in their ground state. 
The reason for this is that only for these fast oscillators an approximate ground state can be found. Indeed, the 
whole bath admits arbitrarily small energy scales which forbid the use of perturbation theory. However, by concentrating 
on the fast bath modes, the relevant energies have a lower cutoff $e^{-\rmd\ell}\omega_c$. It will turn out 
that the energy scales in $\mathcal V$ are even much smaller. 
Thus, we can treat $\mathcal V$ as a small perturbation and construct a systematic perturbation theory for the ground states of the fast bath modes.
%
%
%

Let us reduce the complexity of the problem before returning to the full two-spin case, by noting that the ``free part'' of the Hamiltonian $\mathcal H_B^>$ 
can be written as a sum $\mathcal H_B^> = \sum_n \mathcal H_{B,n}^>$ with $[\mathcal H_{B,n}^>,\mathcal H_{B,n'}^>] = 0$; thus,
each single spin Hamiltonian can be diagonalized separately. 
This is of course not true for the perturbation $\mathcal V$. Still, we will use, for a moment, a mean-field like approach
where $\mathcal V$ is assumed to factorize. With this assumption the summand in $\mathcal V$ 
depending on the $n$-th spin reads
\begin{align}\label{eq:SBmeanfield}
\mathcal V_n &= \frac{h}{2}\left(e^{v_n}\sigma_n^+ + e^{-v_n}\sigma_n^-\right) + \mathcal H_B^<
\nonumber\\ 
&+ \sum_{i_n,<} \lambda_{i_n}\left(a^\dagger_{i_n} + a_{i_n}\right) \sigma_n^z + \frac{Jcm}{2}\sigma_n^z\;,
\end{align}
 where $c$ is the connectivity of the underlying spin model. Upon singling out each spin we treated the surrounding magnetization $m$ as a constant quantum number, i.e. we set $\sigma^z_{n'}/2 = m$ ($n' \ne n$) fixed for $\mathcal V_n$. Since in the following subsection we only deal with single spin Hamiltonians we henceforth omit the index $n$.
Note that the resulting single-spin Hamiltonian is equivalent to (\ref{eq:spinboson}) with $\epsilon = Jcm$ 
 abundantly studied in the literature. 

\section{Renormalization of the single spin-boson system} 

The perturbation associated to the single-spin Hamiltonian becomes
\begin{eqnarray}\label{eq:hamV}
 \mathcal V &=&
 \frac{h}{2}\left(e^{v}\sigma^+ + e^{-v}\sigma^-\right)
 +  \mathcal H_{B}^< 
  \nonumber\\
  && + \sum_{i,<}\lambda_i \left(a^\dagger_i+ a_i \right)\sigma^z 
+ \frac{\epsilon}{2}\sigma^z  
 \; .
\end{eqnarray}

To put the renormalization scheme which we have qualitatively described in the previous paragraph on 
more quantitative grounds we write the full single spin Hamiltonian restricted to the $\vec n_> = 0$ Hilbert space as 
\begin{equation}\label{eq:hamsingle2}
  \mathcal H  = \mathcal H_{B}^> + \mathcal V = 
  \sum_{\vec n_<,s=\pm}\mathcal E_{s,\vec n_<,0}|s, \vec n_< ,0\rangle\langle s, \vec n_<,0|\;.
\end{equation}
The eigenstates $|s,\vec n_< ,0\rangle$ and the energies 
$\mathcal E_{s,\vec n_<,0}$ of the full Hamiltonian 
have to be determined perturbatively. If we denote by $|s\rangle|\vec n_<\rangle|\vec n_>\rangle$ the (factorizing) eigenstates of $\mathcal H_{B}^>$,
it is clear that  $|s,\vec n_<,0\rangle$ are different 
from $|s\rangle|\vec n_<\rangle|0\rangle$.

As one might have guessed, the Hamiltonian on the truncated Hilbert space (\ref{eq:hamsingle2}) can be recast in the original form 
\begin{equation}
  \mathcal H = \frac{\epsilon'}{2} \sigma^z + \frac{h'}{2}\sigma^x + \sum_{i,<}\lambda'_{i}\left(a^\dagger_{i} + a_{i}\right)\sigma^z + 
 \mathcal H_B^<
 \; , 
\end{equation}
 with effective couplings $h'$, $\epsilon'$ and $\lambda_i'$. Hence, by constraining the fast modes to their ground state the original 
 Hamiltonian can be described by an effective Hamiltonian of the same form and with renormalized coupling constants.
 
 The usual recursive procedure is now followed. The unitary transformation 
 $U$ is applied to the bath modes with $e^{-\rmd\ell}\omega'_c<\omega<\omega_c'$ where $\omega_c' = e^{-\rmd\ell}\omega_c$. 
By repeating the above analysis one thus arrives at effective coupling constants recursively defined by the precedent RG step. 
To put it in other words,  one constructs flow equations for all coupling constants.  
As regards the couplings to the bath, it turns out to be more convenient to work with $\alpha$ introduced in (\ref{eq:spectral}) instead of the 
$\lambda_i$s. 

The flow of $\omega_c$ can be immediately written down since, from the very construction of the adiabatic renormalization scheme,
we have
\begin{equation}\label{eq:omegaflow}
  \partial_\ell \omega_c = -\omega_c \; .
\end{equation} 
The perturbation theory breaks down as soon as $h/\omega_c \sim 1$. Hence, there are in general two possibilities: Either the renormalization flow has to be stopped at 
some point and the resulting theory is a spin-boson model with renormalized \emph{finite} coupling constants or the renormalization flow can be pursued until infinity (since 
$h$ might decrease, too, during renormalization) in which case the tunneling vanishes and the spin is in the localized regime.

In the following it will be useful to directly work with the $n$-th order \emph{perturbation operator} $V^{(n)}$ 
defined on the subspace $\mathfrak D$ spanned by the unperturbed eigenstates $|s\rangle|\vec n_<\rangle |0\rangle$. $V^{(n)}$ 
can be written in terms of the matrix elements $\mathcal V_{k,k'} = \langle k|\mathcal V|{k'}\rangle$, with $|k\rangle,|{k'}\rangle \in \mathfrak D$; 
in the Appendix we give its explicit expression up to $4$-th order.  

By using Eq.~(\ref{eq:hamsingle2}) the $n$-th order approximation of the full energies $\mathcal E_{s,\vec n_<,0}$  and states 
$|s,\vec n_<, 0\rangle$ are then the eigenvalues (eigenstates) of $V^{(n)}$. The $n$-th order approximation of 
the Hamiltonian restricted to the $\vec n_> = 0$ Hilbert space (\ref{eq:hamsingle2}) can therefore be written in the form
\begin{equation}\label{eq:hamsingle3}
  \mathcal H^{(n)} = V^{(n)} \; ,
\end{equation}  
by noting that the unperturbed energy is $E_{\vec n_> = 0} = 0$ (note that we denote the unperturbed energies by a standard $E$). 
The effective Hamiltonian is thus completely determined by the perturbation operator which we shall calculate in the following subsections.

\subsection{First order adiabatic renormalization}

We use degenerate static perturbation theory (see App.~\ref{sec:rendetails} for details) to calculate $V^{(n)}$. 
The degenerate subspace $\mathfrak D$ is spanned by the states 
$|+\rangle|\vec n_<\rangle|0\rangle$ and $|-\rangle|\vec n_<\rangle|0\rangle$ which have zero unperturbed energy (with respect to the operator $\mathcal H_B^>$). 
The perturbation subsequently splits the degenerate 
energy into different energy levels as usual. 

The first-order perturbation operator, $V^{(1)}$, has matrix elements $V^{(1)}_{k,k'} = \mathcal V_{k,k'}$ with $|k\rangle,|{k'}\rangle \in \mathfrak D$. 
We call diagonal elements all matrix elements which conserve $\vec n_<$. At this order all off-diagonal elements are obviously zero and we can work at fixed $\vec n_<$. Since the fast bath modes act solely on the $e^{\pm v}$-terms in 
$\mathcal V$ only $h$ is modified at this order in perturbation theory. More precisely, in the fixed $\vec n_<$-subspace we have 
\begin{equation}
  V^{(1)} = \left(
    \begin{array}{cc}
      E_{\vec n_<} + \epsilon/2 & h'/2 \\
      h'/2 & E_{\vec n_<} - \epsilon/2 \\
    \end{array}
    \right)\;,
\end{equation}
with $E_{\vec n_<} = \sum_{i,<} \omega_i n_{<,i}$ and
\begin{align}
  h'\equiv h\langle0|e^{\pm v}|0\rangle =  \exp\left[-\sum_{j,>}\frac{2\lambda_j^2}{\omega_j^2}\right]
= h e^{-\alpha\omega_c^{s-1}\rmd\ell} \; .
\end{align}
In the last step we used the well-known identity $e^{-2r[a^\dagger_j - a_j]} = e^{-2r^2} e^{-2r a_j^\dagger}e^{2r a_j}$ (for $r \in \mathbb{R}$) and the definition of $\alpha$ [see (\ref{eq:spectral})]: $\sum_{j,>}2\lambda_j^2/\omega_j^2 = \alpha \int_{e^{-\rmd\ell}\omega_c}^{\omega_c} \ \rmd\omega \ \omega^{s-2} = \alpha \omega_c^{s-1}\rmd\ell$.

Until here the adiabatic renormalization scheme is only a more formal presentation than the ``adiabatic renormalization'' previously used for an Ohmic bath in~\cite{CalLegg87,We08} to calculate
the renormalized tunneling 
amplitude $h'$. At each renormalization step $h$ is diminished by the Franck-Condon factor $e^{-\alpha\omega_c^{s-1}\rmd\ell}$ and the total decrease after 
$\ell /\rmd\ell$ such steps is $h(\ell) = e^{-\alpha\omega_c^{s-1}\ell}h$. The 
flow has to be interrupted as soon as $\omega_c(\ell) = e^{-\ell}\omega_c$ becomes of the order of $h(\ell)$. This condition can be reformulated in a 
more concise way: the variable 
\begin{equation}
  \tilde h = \frac{\sqrt{2}h}{\omega_c}
\end{equation} 
has to remain small during renormalization (the numerical factor in the definition of $\tilde h$ is introduced for later convenience) and, by using the previous equations, it scales as 
\begin{equation}\label{eq:RGh1}
\partial_\ell \tilde h = (1 - \tilde\alpha) \tilde h \; ,
\end{equation}
where 
\begin{equation}
  \tilde\alpha = \alpha \omega_c^{s-1}\;. 
\end{equation}
In the Ohmic case, this equation reduces to (\ref{eq:RG11}) and 
predicts the localization transition since for $\alpha>1$ the procedure can be 
infinitely repeated and $\tilde h$ therefore scales to zero. For $\alpha<1$ the flow has to be stopped as soon as $\tilde h \sim 1$.  

However, the adiabatic renormalization scheme can be carried out beyond the first order calculation. In the following subsections we shall determine the perturbation operator up to third order to find the RG equations for $J$, $\epsilon$, $\tilde h$ and $\tilde\alpha$. 

\subsection{Higher order result: Renormalization of $\tilde h$ and $\epsilon$.}

In this subsection we use  the perturbative results detailed in App.~\ref{sec:rendetails}. 
We first focus on the elements diagonal in $\vec n_<$. By using Eq.~(\ref{eq:Vthird}) we have 
for the third order perturbation operator in the $\pm$-basis (at fixed $\vec n_<$)
\begin{align}
\label{eq:corrmatrix}
  &V ^{(3)} = \\
&\left(
    \begin{array}{cc}
      E_{\vec n_<} + \frac{\epsilon}{2} + \mathcal V_{+,+}^{(2)} + \mathcal V_{+,+}^{(3)}  
& \frac{h'}{2} + \mathcal V_{+,-}^{(3)} \\
      \frac{h'}{2} + \mathcal V_{-,+}^{(3)} & 
E_{\vec n_<} -\frac{\epsilon}{2} + \mathcal V_{-,-}^{(2)} + \mathcal V_{-,-}^{(3)}\\
    \end{array}
  \right)
  \nonumber
  \;,
\end{align}
with 
\begin{equation}\label{eq:V2}
  \mathcal V^{(2)}_{s,s'} = \sum_{k_1 \notin\mathfrak D} 
\frac{\langle s|\langle\vec n_<|\langle 0|\mathcal V|k_1\rangle\langle k_1|\mathcal  V |s'\rangle|\vec n_<\rangle|0\rangle}{-E_{k_1}}
\end{equation}
and
\begin{widetext}
\begin{eqnarray}\label{eq:V3}
  \mathcal V^{(3)}_{s,s'} = \sum_{\substack{k_1 \notin\mathfrak D \\ k_2 \notin\mathfrak D}} 
\frac{\langle s|\langle\vec n_<|\langle 0|\mathcal V|k_1\rangle\langle k_1|\mathcal V|k_2\rangle\langle k_2|\mathcal  V |s'\rangle|\vec n_<\rangle|0\rangle}{E_{k_1}E_{k_2}}
- \sum_{\substack{k_1\notin\mathfrak D\\k_2\in\mathfrak D}}\frac{\langle s|\langle\vec n_<|\langle 0|\mathcal V|k_1\rangle\langle k_1|\mathcal V|k_2\rangle\langle k_2|\mathcal  V |s'\rangle|\vec n_<\rangle|0\rangle}{E_{k_1}^2}\; , 
\end{eqnarray}
\end{widetext}
where $E_{k_1}$ is the (unperturbed) energy of the (unperturbed) state $|k_1\rangle$. 

Let us now determine the action of each summand in $\mathcal V$ given in Eq.~(\ref{eq:hamV}) when inserted into Eqs.~(\ref{eq:V2}) and (\ref{eq:V3}). 
First of all, the $\sum_{i,<}\lambda_i(a^\dagger_i + a_i)\sigma^z$-term does not contribute. Indeed, for fixed $\vec n_<$ such a term would 
have to occur twice to give a non-zero contribution. But in this case, the matrix elements in (\ref{eq:V2}) and (\ref{eq:V3}) would give identically 
zero since $\mathfrak D$ is invariant under $\sum_{i,<}\lambda_i(a^\dagger_i + a_i)\sigma^z$ and the sum runs over $k_1 \ne \mathfrak D$. 
The same is true for the $\sum_{i,<}\omega_ia^\dagger_i a_i$-term when considering $\mathcal V^{(2)}$. $\mathfrak D$ is invariant under its action and it therefore cannot contribute in (\ref{eq:V2}). 

We realize that $\mathcal H_B^<$ could yield a finite contribution  if inserted for $\mathcal V$ into the second factor of the first summand and into the third factor of the second summand of $\mathcal V^{(3)}$. For the first summand this requires $k_1 = k_2$ and for the second summand $k_2 = |s'\rangle|\vec n_<\rangle| 0\rangle$. Obviously, these two summands then cancel exactly.

All relevant diagonal (with respect to $\vec n_<$) contributions thus stem from the terms proportional to $\epsilon$ and $h$ in $\mathcal V$.
Note that $\mathcal V$ always induces a spin flip if we consider only the term proportional to $h$. 
Since the $\epsilon \sigma^z/2$-term leaves $\mathfrak D$ invariant (and hence cannot be inserted into $\mathcal V^{(2)}$), $\mathcal V^{(2)}$ can only contribute matrix elements with no spin flip. Moreover, this matrix element is not proportional to any other operator, since it turns out that $\mathcal V_{+,+}^{(2)}=\mathcal V_{-,-}^{(2)}$: Therefore it contributes only an irrelevant constant energy. 

The more interesting contribution comes from $\mathcal V^{(3)}$.
The elements of $\mathcal V^{(3)}$ with one spin flip read [see (\ref{eq:V3})]:
\begin{widetext}
\begin{align}\label{eq:V31}
  \mathcal V^{(3)}_{+,-} &= 
\frac{h^3}{8}\sum_{k_1,k_2\ne\mathfrak D}
\frac{\langle + |\langle\vec n_<|\langle 0| e^{v}\sigma^+|k_1\rangle \langle k_1|e^{-v}\sigma^- | k_2 \rangle  
\langle k_2 | e^{v}\sigma^+ |-\rangle|\vec n_<\rangle|0\rangle}{E_{k_1}E_{k_2}}\nonumber\\
&-\frac{h^3}{8}\sum_{k_1\ne\mathfrak D}
\frac{\langle + |\langle\vec n_<|\langle 0| e^{v}\sigma^+|k_1\rangle \langle k_1|e^{- v}\sigma^-| 0 \rangle  
\langle 0 | e^{v}\sigma^+ |-\rangle|\vec n_<\rangle|0\rangle}{E_{k_1}^2}\nonumber\\
&= \frac{h^3}{8}\sum_j \frac{\langle 0 |e^{v}| j\rangle\langle j |e^{-v}| j \rangle\langle j |e^{v} |0\rangle}{\omega_j^2} 
- \frac{h^3}{8}\sum_j \frac{\langle 0 |e^{v}| j\rangle\langle j |e^{-v}| 0 \rangle\langle 0 |e^{v} |0\rangle}{\omega_j^2} + \cdots \;,
\end{align}
\end{widetext}
where $|j\rangle$ is the oscillator state with all oscillators except the $j$-th one in the ground state, and the $j$-th oscillator in its first excited state. Note that the terms in (\ref{eq:V31}) are more and more suppressed by the factor $\lambda_j/\omega_j \sim \lambda_j/\omega_c$ and we will neglect all the terms not explicitly listed in (\ref{eq:V31}). 
After a straightforward calculation one has 
\begin{eqnarray*}
\langle 0 |e^{\pm v} | j\rangle &=& \mp e^{-\tilde\alpha \rmd \ell} \frac{2\lambda_j}{\omega_j} + \cdots\;,\\
\langle j |e^{\pm v} | 0\rangle &=& \pm e^{-\tilde\alpha \rmd \ell} \frac{2\lambda_j}{\omega_j} + \cdots\;.
\end{eqnarray*}
Therefore we find by using $\omega_j \simeq \omega_{j'} \simeq \omega_c$ 
\begin{eqnarray}\label{eq:V32}
  \mathcal V^{(3)}_{+,-} &=& -{h'}^3\sum_j\frac{\lambda_j^2}{\omega_j^4} 
+ \mathcal O(\omega_c^{-4}) \; .
\end{eqnarray}
In conjunction with $\sum_{j,>}\frac{2\lambda_j^2}{\omega_j^4} = \alpha \omega_c^{s-3}\rmd\ell$ this leads to
\begin{eqnarray}\label{eq:V32}
  \mathcal V^{(3)}_{+,-} &=& -\frac{{h'}^3}{2\omega_c^2}\alpha\omega_c^{s-1}\rmd\ell + \mathcal O(\rmd\ell^3) .
\end{eqnarray}
From the form of (\ref{eq:corrmatrix}) we directly deduce the field $h'$ which now scales as
\begin{equation}\label{eq:scalingh}
  h' = h \exp\left[-\tilde\alpha\rmd\ell - \frac{{\tilde h}^2}{2}\tilde\alpha\rmd\ell + \mathcal O(\omega_c^{-3},\rmd\ell^3)\right]\; . 
\end{equation}
The final flow equation reads
\begin{eqnarray}\label{eq:RGh22}
  \partial_\ell \tilde h &=& \left(1 - \tilde\alpha - \frac{{\tilde h}^2\tilde\alpha}{2} \right) \tilde h + \mathcal O(\tilde h^4)
\end{eqnarray}
and this equation extends (\ref{eq:RGh1}) to the next leading order in $\tilde h$.

We now analyze the flow of the bias $\epsilon$. By inserting $\epsilon \sigma^z/2$ into $\mathcal V^{(3)}$ 
once in each summand we find the elements with no spin flip:
\begin{widetext}
\begin{align}
  \mathcal V^{(3)}_{+,+} &= 
\frac{h^2\epsilon}{8}\sum_{k_1,k_2 \notin\mathfrak D} 
\frac{\langle +|\langle \vec n_<|\langle 0|e^v\sigma^+|k_1\rangle\langle k_1|\sigma^z|k_2\rangle\langle k_2|e^{-v}\sigma_- |+\rangle|\vec n_<\rangle|0\rangle}{E_{k_1}E_{k_2}} \\ 
&- \frac{h^2\epsilon}{8}\sum_{\substack{k_1\notin\mathfrak D\\k_2\in\mathfrak D}}\frac{\langle +|\langle \vec n_<|\langle 0|e^v\sigma^+|k_1\rangle\langle k_1|e^{-v}\sigma^-|k_2\rangle\langle k_2|\sigma^z|+\rangle|\vec n_<\rangle|0\rangle}{E_{k_1}^2}\; . \nonumber
\end{align}
\end{widetext}
It is easy to show that $\mathcal V^{(3)}_{-,-} = -\mathcal V^{(3)}_{+,+}$.
The explicit expression for $\mathcal V_{+,+}^{(3)}$ is given by 
\begin{eqnarray}
  \mathcal V_{+,+}^{(3)} &=& -\frac{h^2\epsilon}{8}\sum_j\frac{\langle0|e^{v}|j\rangle
\langle j|j\rangle\langle j|e^{-v}|0\rangle}{\omega_j^2} \nonumber\\
&-&\frac{h^2\epsilon}{8}\sum_j\frac{\langle0|e^{v}|j\rangle
\langle j|e^{-v}|0\rangle\langle 0|0\rangle}{\omega_j^2} \nonumber\\
&=& -\frac{{h'}^2\epsilon}{2\omega_c^2}\tilde\alpha\rmd\ell \; .
\end{eqnarray}
Consequently, the bias $\epsilon$ is renormalized according to
\begin{equation}\label{eq:RGepsi}
  \partial_\ell \epsilon = - \frac{\tilde h^2\tilde\alpha}{2} \epsilon \; .
\end{equation}
The above equation has been derived the first time in~\cite{Chakravarty82} by using the mapping to the classical long-range Ising chain. The variation of $\epsilon$ is of second order in $\tilde h$ and therefore small. Hence, the ratio $\epsilon/\omega_c$ does not remain small during the adiabatic renormalization. In order to obtain a complete picture of the transition gouverned by (\ref{eq:RGh22}), (\ref{eq:RGalpha}) and (\ref{eq:RGepsi}) it is therefore necessary to study the regime $\epsilon \gg h$, as well. This will be done in Sec.~\ref{sec:largeepsi}.

\subsection{Renormalization of $\alpha$}

We now turn to the ``off-diagonal'' elements of $V^{(3)}$ which induce a change in the quantum numbers $\vec n_<$, i.e. matrix elements proportional to $\sum_{i,<}\lambda_i(a^\dagger_i + a_i)\sigma^z$. Again, this term leaves the subspace $\mathfrak D$ invariant, and it cannot occur in $\mathcal V^{(2)}$. But we can insert it for $\mathcal V$ into the second factor of the first summand and for $\mathcal V$ into the third factor of the second summand of $\mathcal V^{(3)}$. This time the two summands do \emph{not} cancel. Indeed, the $\sigma^z$ creates the necessary minus sign to yield a finite contribution as we have already seen when we discussed the renormalization of $\epsilon$. By repeating the analysis that led us to (\ref{eq:RGepsi}) we find a similar RG equation:
\begin{equation}\label{eq:RGlambda}
  \partial_\ell \lambda_i = - \frac{\tilde h^2\tilde\alpha}{2} \lambda_i \; .
\end{equation} 
Since the overall amplitude of the $\lambda_i$ scale as $\lambda^2_i \sim \alpha$ we can write the equivalent equation
\begin{equation}\label{eq:RGalpha}
  \partial_\ell \tilde \alpha = (1-s)\tilde\alpha -\tilde h^2 \tilde \alpha^2  \; ,
\end{equation}
where the first term of the rhs in the above equation comes from the purely dimensional scaling of the $\omega_c^{s-1}$-prefactor of $\tilde \alpha$ [see (\ref{eq:omegaflow})].

These RG equations are equivalent in the limit $\tilde h \to 0$ to the ones derived by Kosterlitz~\cite{Kosterlitz76} in the vicinity of the fixed point $\{ \tilde\alpha=1,\tilde h = \sqrt{1-s} \}$. 
Note that the condition of validity $\tilde h \ll 1$ limits the use of the RG equations to $1-s \ll 1$~\cite{Vojta05} for $s\le1$.    

We have thus derived the full RG equations of the biased spin-boson model within a systematic adiabatic renormalization scheme, without using the cumbersome mapping to a long-range Ising chain and its subsequent renormalization.

\subsection{The large-$\epsilon$ regime}\label{sec:largeepsi}

When $\epsilon(\ell)/h(\ell)$ becomes large under its flow equation it is necessary to modify the adiabatic RG we have used so far to account for the fact that $\epsilon(\ell)$ might be of the same order as $\omega_c(\ell)$ after some RG steps. The perturbation associated to the single-spin case reads now 
\begin{equation}
  \mathcal V = \frac{h}{2}\left(e^v \sigma^+ + e^{-v}\sigma^-\right) + \mathcal H_B^< + \sum_{i,<}\lambda_i\left(a_i^\dagger + a_i\right)\sigma^z \; ,
\end{equation}
since we have to include the bias term into the unperturbed part of the Hamiltonian,
\begin{equation}
  \mathcal H^0 = \mathcal H_B^> + \frac{\epsilon}{2}\sigma^z \; ,
\end{equation}
which groups all high-energy terms together. The resulting perturbation series will be performed with respect to the small ratios $h/\omega_c$ and $h/\epsilon$. We first analyze the renormalization of $\alpha$ in the large $\epsilon$ case. Note that, in contrast to the previous calculation, the $s = \pm$-states now lie on \emph{different} degenerate subspaces. For $\epsilon > 0$ the lowest energy state is $|-\rangle|\vec n_<\rangle|0\rangle$. The analysis of the previous subsection can now be repeated to yield
\begin{align}\label{eq:lambdaepsi}
  \lambda'_i - \lambda_i &= -\frac{{h'}^2\lambda_i}{4}\sum_j\frac{\langle0|e^{v}|j\rangle
\langle j|j\rangle\langle j|e^{-v}|0\rangle}{(\epsilon + \omega_j)^2} 
- \frac{{h'}^2\lambda_i}{4\epsilon^2}
\nonumber\\
& - \frac{{h'}^2\lambda_i}{4}\sum_j\frac{\langle0|e^{v}|j\rangle
\langle j|e^{-v}|0\rangle\langle 0|0\rangle}{(\epsilon + \omega_j)^2} - \frac{{h'}^2\lambda_i}{4\epsilon^2}
\nonumber \\
&= - \frac{{h'}^2\lambda_i}{2\epsilon^2} - \frac{{h}^2\lambda_i}{(\epsilon + \omega_j)^2}\tilde\alpha\rmd\ell\; ,
\end{align}
where $\lambda_i'$ is the effective bath coupling up to second order in $h$ which is essentially equal to $\lambda_i$ minus an offset term $h^2\lambda_i/2\epsilon^2$ which stems from the energy asymmetry of the two spin states (due to $\epsilon>0$). 
Indeed, let us rotate the spin in (\ref{eq:spinboson}) around the $y$-axis with an angle $\theta$ given by $\tan 2\theta = h/\epsilon$. We then find the transformed Hamiltonian
\begin{widetext}
\begin{equation}\label{eq:sb2}
  \mathcal H'_{\rm SB} = \frac{1}{2}\sqrt{h^2+\epsilon^2} \ \sigma^z + \sigma^z\sum_i\lambda'_i(a_i^\dagger + a_i) 
- \frac{h}{\epsilon}\sigma^x\sum_i\lambda'_i(a_i^\dagger + a_i) + \mathcal H_B \; ,
\end{equation}
\end{widetext}
where we defined $\lambda_i' \equiv \epsilon\lambda_i/\sqrt{h^2+\epsilon^2}
\simeq \lambda_i\left[1 - h^2/2\epsilon^2\right]$. The perturbation series in (\ref{eq:lambdaepsi}) thus gives back the correct effective $\lambda_i'$.

The terms proportional to $\rmd\ell$ come from the high energy bath modes and they lead to the renormalization of $\tilde\alpha$, the flow equation of which is found to be
\begin{equation}\label{eq:RGaepsi}
  \partial_\ell\tilde\alpha = \left(1-s\right)\tilde\alpha 
+ \frac{\tilde h^2}{\tilde \epsilon^2}\tilde\alpha^2- \frac{\tilde h^2}{(1 + \tilde\epsilon)^2}\tilde\alpha^2 \; ,
\end{equation}
with $\tilde\epsilon \equiv \epsilon/\omega_c$. Note that we used $h' = (1-\tilde\alpha\rmd\ell)h$ by making the underlying assumption that $h$ is not renormalized. This will be discussed in the following paragraph.

The derivation of the effective tunneling is more subtle since it involves a matrix element between two distinct adiabatic subspaces (note that the $s = \pm$-states have different energies). In the present single-spin case the adiabatic renormalization scheme cannot find the renormalization of both $\epsilon$ and $h$: For fixed $\vec n_<$ the degenerate subspace $\mathfrak D_-$ associated to $|-\rangle|\vec n_<\rangle|0\rangle$ is one-dimensional, thus delivering only one renormalization equation instead of two needed to determine the RG equations of $\epsilon$ and $h$. By using the transformed Hamiltonian (\ref{eq:sb2}) we have 
\begin{equation}\label{eq:RGE}
  \partial_\ell \frac{1}{2}\sqrt{\epsilon^2 + h^2} = \frac{h^2}{\epsilon} \frac{\tilde\alpha}{\tilde\epsilon(\tilde\epsilon+1)} \; .
\end{equation}
For large $\tilde\epsilon$ the renormalization flow is cut off and the ground state energy remains constant. The mapping of the biased spin-boson model to a classical long-range interacting Ising chain informs us that no phase transition occurs (the bias translates into a magnetic field \emph{via} the mapping). Such a conclusion is totally compatible with (\ref{eq:RGE}). In the following section we show that the adiabatic renormalization scheme allows -- in the two-spin case -- to find the explicit RG equations for both $h$ and $J$ in contrast to the present single-spin case.

\section{Two coupled spins: Renormalization for small inter-spin coupling}
\label{sec:2spins}

We now come back to the two-spin Hamiltonian (\ref{eq:ham2s}). In contrast to the analysis presented in the previous section, the Hilbert space of the degenerate states $\mathfrak D$ for fixed $\vec n_<$ has now four dimensions spanned by the four eigenstates of the $\sigma_1^z\sigma_2^z$-operator $|+_1\rangle|+_2\rangle$, $|+_1\rangle|-_2\rangle$, $|-_1\rangle|+_2\rangle$ and $|-_1\rangle|-_2\rangle$. 

In order to write down the perturbation operator we remark that $\mathcal V^{(3)}$ changes $|+_1\rangle|+_2\rangle$ into $|+_1\rangle|-_2\rangle$ or $|-_1\rangle|+_2\rangle$ into $|+_1\rangle|+_2\rangle$
etc., if only the term proportional to $h$ is inserted. It leaves the spin-state invariant if $J\sigma_1^z\sigma_2^z/4$ is inserted once. Hence, $\mathcal V^{(3)}$ has matrix elements with \emph{one} spin flip and \emph{zero} spin flip. 
The case of $\mathcal V^{(2)}$ is a bit more complicated, since it could have matrix elements with \emph{a priori} \emph{no} spin flip and ones with \emph{two} spin flips. It is easy to show that the matrix elements with two spin flips have to vanish since the two baths are uncorrelated. 
Accordingly, we denote the matrix elements of $\mathcal V^{(3)}$ with one spin flip (which are all equal) by $\mathcal V_1^{(3)}$ and the matrix elements with no spin flip by $\mathcal V^{(2)}_0$ and $\mathcal V_0^{(3)}$, respectively. No finite matrix element corresponding to two spin flips arises in the perturbation operator up to third order. It can be shown that, since $\mathfrak D$ is invariant under a double spin-flip, no matrix element with two spin-flips can be generated under the RG flow to \emph{any} order. We have explicitly verified that the fourth order contributions to a double spin-flip matrix element cancel. For details we refer the reader to the appendix.

Let us now discuss the RG equations of $J$, $\tilde\alpha$ and $\tilde h$ for the two-spin case. We use the third order perturbation operator in the four dimensional diagonal ($\vec n_<$ fixed) degenerate subspace $\mathfrak D$:
\begin{widetext}
\begin{equation}\label{eq:fullV}
  V^{(3)} = \left(
    \begin{array}{cccc}
      J/4 + \mathcal V_0^{(2)} - \mathcal V_0^{(3)} & h'/2 + \mathcal V_1^{(3)} & h'/2 + \mathcal V_1^{(3)} & 0  \\
    h'/2 + \mathcal V_1^{(3)} & -J/4 + \mathcal V_0^{(2)} +\mathcal V_0^{(3)}& 0 & h'/2 + \mathcal V_1^{(3)} \\
    h'/2 + \mathcal V_1^{(3)} & 0 & -J/4 + \mathcal V_0^{(2)} +\mathcal V_0^{(3)}& h'/2 + \mathcal V_1^{(3)} \\
   0  &  h'/2 + \mathcal V_1^{(3)} & h'/2 + \mathcal V_1^{(3)} & J/4 + \mathcal V_0^{(2)} -\mathcal V_0^{(3)}\\
   \end{array}
   \right) \; ,
\end{equation}
\end{widetext}
where we omitted to explicitly write down $E_{\vec n_<}$ to clear up the notations. Let us first discuss the inter-spin coupling $J$.
The effective coupling in the $z$-direction is renormalized in a similar way as the bias in the single-spin case: Since the term $\mathcal V_0^{(2)}$ has no alternating sign it does not contribute to the $J\sigma_1^z\sigma_2^z/4$-part of the Hamiltonian but rather to an irrelevant total energy shift. It is straightforward to show that 
\begin{equation}
  \mathcal V_0^{(3)} = 2\mathcal V_{++}^{(2)} \;,
\end{equation}
where the factor $2$ comes from the fact that the $e^{\pm v}$-terms can be inserted for each of the two baths.
The RG equation for $J$ is hence given by
\begin{equation}\label{eq:RGJ}
  \partial_\ell J = -\frac{\tilde \alpha(\sqrt{2}\tilde h)^2}{2} J \; .
\end{equation}
The magnetic field $\tilde h$ and the bath strength verify the equations
\begin{align}\label{eq:RGJha}
  \partial_\ell \sqrt{2}\tilde h &= \left(1-\tilde\alpha-\frac{\tilde \alpha (\sqrt{2}\tilde h)^2}{2}
\right)\sqrt{2}\tilde h \; ,\\
\partial\tilde\alpha &= (1-s)\tilde\alpha - \tilde\alpha^2(\sqrt{2}\tilde h)^2 \; .
\end{align}
By absorbing the factor $\sqrt{2}$ again in a redefinition of $\tilde h$ one arrives at the same equations as for the single-spin case, with the only difference that $J$ plays the role of $\epsilon$.

\subsection{Analysis for large $J$.}\label{sec:largeJ}

Very much as for the single-spin case the perturbation series breaks down as soon as $J(\ell)/h(\ell)$ becomes large. Hence, we have to generalize the analysis presented in Sec.~\ref{sec:largeepsi} for two spins in order to gain an insight into the RG flow when $J(\ell)/h(\ell) \gg 1$. 

\subsubsection{The case $\tilde\alpha = 0$.}

We first discuss the case where the bath is decoupled from the system. The lowest energy subspace $\mathfrak D_-$ is spanned by the two eigenvectors $|+-\rangle$ and $|-+\rangle$ where we used the short-hand notation $|-+\rangle \equiv |-_1\rangle|\vec n_{1,<}\rangle|0_1\rangle|+_2\rangle|\vec n_{2,<}\rangle|0_2\rangle$ with an analogue definition for $|+-\rangle$. We assumed here -- without loss of generality -- that $J>0$. A finite $h$ leads to an energy splitting within $\mathfrak D_-$ in contrast to the single-spin case. Indeed, if we start from the two-spin Hamiltonian~(\ref{eq:ham2s}) without dissipation, the four energy levels of the Hamiltonian are easily found to be $\mathcal E_1 = -J/4$, $\mathcal E_2 = J/4$, $\mathcal E_3 = -\sqrt{J^2 + 16h^2}/4$ and $\mathcal E_4 = \sqrt{J^2 + 16h^2}/4$. 
The high-energy bath modes modify the two energies $\mathcal E_1$ and $\mathcal E_3$ which we will analyse separately to obtain two RG equations for the two couplings $J$ and $h$. In contrast to the single-spin case it is thus possible to obtain RG equations \emph{via} the adiabatic RG scheme for \emph{all} parameters by restricting the system to $\mathfrak D_-$. 

Note that in contrast to the analysis in Sec.~\ref{sec:2spins} a double-spin flip matrix element occurs at second order in $h/J$ since
\begin{align}\label{eq:V22large}
  \mathcal V^{(2)}_2 &= \frac{h^2}{4}\sum_{k_1\notin\mathfrak D_-}\frac{\langle -+|\mathcal V|k_1\rangle\langle k_1|\mathcal V|+-\rangle}{-J/2} \nonumber\\
&= -\frac{{h}^2}{J} = \mathcal V_0^{(2)} \; .
\end{align}
This result is expected since the symmetry argument presented in the previous section fails: 
$\mathfrak D_-$ is not invariant under two spin-flips. The last equation of the rhs of (\ref{eq:V22large}) shows that at second order the two spin-flip element is equal to the zero spin-flip element. Let us write the two-spin Hamiltonian without dissipation restricted to $\mathfrak D_-$ as
\begin{equation}
  \mathcal H_- = \mathcal E_1|\psi_1\rangle\langle\psi_1| + \mathcal E_3|\psi_3\rangle\langle\psi_3| \; .
\end{equation}  
The second order perturbation matrix associated to $\mathfrak D_-$ reads
\begin{equation}
  V^{(2)}_- = \left(\begin{array}{cc}
     -h^2/J & -h^2/J \\
    -h^2/J & -h^2/J \\
    \end{array}\right)\;,
\end{equation}
where we used (\ref{eq:V22large}). It is then straightforward to find the perturbative energies and eigenstates
\begin{align}\label{eq:eigen}
  \mathcal E_1  &\simeq -\frac{J}{4}\;\;\mathrm{and}\;\;|\psi_1\rangle \simeq (-1,1)\;, \\
  \mathcal E_3  &\simeq -\frac{J}{4}-\frac{2h^2}{J}\;\;\mathrm{and}\;\;|\psi_3\rangle \simeq (1,1)\;,
\end{align}
which are obviously equal up to second order to the exact states and energies introduced above.

\subsubsection{The dissipative case.}

When the bath is coupled to our system we have to modify the result (\ref{eq:V22large}). Indeed, the sum over the intermediate states $|k_1\rangle$ now also runs over the excited states of the high-energy bath modes and
\begin{equation}
  V_-^{(2)}= \left(\begin{array}{cc}
    -\frac{{h'}^2}{J} - \frac{h^2\tilde\alpha\rmd\ell}{\omega_c + J/2} & -\frac{{h'}^2}{J} \\
    -\frac{{h'}^2}{J} & -\frac{{h'}^2}{J} - \frac{h^2\tilde\alpha\rmd\ell}{\omega_c + J/2} \\
    \end{array}\right)\;.
\end{equation} 
It follows that in the fixed $\vec n_<$ subspace the two eigenvalues read
\begin{align}\label{eq:RGeigen}
  \mathcal E_1  &\simeq -\frac{J}{4} - \frac{h^2\tilde\alpha\rmd\ell}{\omega_c + J/2}\;, \\
  \mathcal E_3  &\simeq -\frac{J}{4} - \frac{h^2\tilde\alpha\rmd\ell}{\omega_c + J/2}- \frac{2{h'}^2}{J}\;,
\end{align}
from which we infer the RG equations
\begin{align}\label{eq:RHJh}
  \partial_\ell \tilde J &= \tilde J + \frac{\tilde h^2\tilde\alpha}{1 + \tilde J/2} \; ,\nonumber\\
  \partial_\ell \tilde h &= (1-\tilde\alpha)\tilde h \; ,
\end{align}
with $\tilde J = J/\omega_c$. Before discussing these equations let us analyse the renormalization of $\tilde\alpha$ by employing the same method as in Sec.~\ref{sec:largeepsi}. The only difference is that the \emph{two} terms $e^{\pm v_{1,2}}\sigma_{1,2}^\pm$ can now be inserted for $\mathcal V$ into $\mathcal V^{(3)}$. To be more precise, the renormalization of the $\sigma_1^z\lambda_i(a^\dagger_i + a_i)$-term is given by
\begin{align}
  \mathcal V_{+-,+-,\lambda_i}^{(3)} &= -\frac{h^2}{4}\sum_{j_1}\frac{\langle0|e^{v_1}|j_1\rangle
\langle j_1|-\lambda_i|j_1\rangle\langle j_1|e^{-v_1}|0\rangle}{(J/2+\omega_{j_1})^2} \nonumber\\
&-\frac{h^2}{4}\sum_{j_2}\frac{\langle0|e^{-v_2}|j_2\rangle
\langle j_2|\lambda_i|j_2\rangle\langle j_2|e^{v_2}|0\rangle}{(J/2+\omega_{j_2})^2} \nonumber\\
&-\frac{h^2}{4}\sum_{j_1}\frac{\langle0|e^{v_1}|j_1\rangle
\langle j_1|e^{-v_1}|0\rangle\langle 0|\lambda_i|0\rangle}{(J/2+\omega_{j_1})^2} \nonumber\\
&-\frac{h^2}{4}\sum_{j_2}\frac{\langle0|e^{v_2}|j_2\rangle
\langle j_2|e^{-v_2}|0\rangle\langle 0|\lambda_i|0\rangle}{(J/2+\omega_{j_2})^2} \nonumber\\
&- \frac{{h'}^2}{2(J/2)^2} 
= -\frac{2{h'}^2\lambda_i}{J^2} -\frac{4{h}^2\lambda_i}{(J+2\omega_c)^2}\tilde\alpha\rmd\ell \; ,
\end{align}
where we did not explicitly write down the occupation numbers of the slow-bath modes. Note that the two first lines cancel. Here, $h'$ is according to (\ref{eq:RHJh}) the effective magnetic field after one RG step and therefore 
\begin{equation}\label{eq:RGJa}
  \partial\tilde\alpha = \left(1-s\right)\tilde\alpha - \frac{\tilde h^2\tilde\alpha^2}{(1 + 2\tilde J)^2} \; ,
\end{equation}
where -- again -- $\tilde h = 2h/\omega_c$.

Eq.~(\ref{eq:RGJa}) is very different from its counterpart (\ref{eq:RGalpha}) of the unbiased spin-boson system. Indeed, the $\tilde J$ in the denominator becomes large when $\omega_c$ decreases such that in this limiting case the effective RG flow reads $\partial_\ell\tilde\alpha \simeq (1-s)\tilde\alpha$. Let us first discuss the Ohmic case for which $s = 1$. The Kosterlitz--Thouless phase transition is then clearly destroyed and replaced by a second order phase transition for $\tilde h$. After the RG flow of $\tilde\alpha$ in the small $J$ regime [see (\ref{eq:RGJha})] which is Kosterlitz-Thouless like, the flow is cut off at large $J$ according to (\ref{eq:RGJa}). The bath strength has then attained some effective value $\alpha^*$ which remains essentially constant as soon as $\tilde J$ is large. Depending on whether $\alpha^*>1$ or $\alpha^*<1$ the spins are localized or delocalized. 

For a super-Ohmic spectral density the spins are always delocalized since $\tilde\alpha\to 0$ in this case, very much as in the single-spin system. However, for a \emph{sub-Ohmic} bath (\ref{eq:RGJa}) predicts an ever growing $\tilde\alpha$ so that $\tilde h$ renormalizes \emph{always} to zero. This is in sharp contrast to the Kosterlitz flow of the sub-Ohmic single spin-boson system which predicts a second order phase transition~\cite{Vojta05}.

\section{Conclusion}

We have shown that the higher-order adiabatic renormalization scheme is able to reproduce the renormalization group (RG) equations for a single spin-boson system. For an unbiased system, our RG equations (\ref{eq:RGh22}) and (\ref{eq:RGalpha}) are equivalent to the ones derived by Kosterlitz for the long-range Ising chain~\cite{CalLegg87,Kosterlitz76} in the limit $h/\omega_c \ll 1$. 
When our method is applied to the biased spin-boson system, the adiabatic RG fails to predict an RG equation for each parameter since the relevant subspace, in which the adiabatic RG operates, is too small for such a purpose, in contrast to the more interesting noisy two-spin system for which the RG flow of each parameter can be found. At $\alpha = 1/2$ the so-called coherent-incoherent crossover takes place. The adiabatic RG fails to capture this transition since $\tilde h$ grows strongly for $\alpha = 1/2$ leading to a breakdown of perturbation theory. To properly analyse the coherent-incoherent transition one probably needs to first transform the spin-boson Hamiltonian~\ref{eq:spinboson} into the so-called ``resonance level model'' [see~\cite{CalLegg87} for details] which is a free theory for $\alpha = 1/2$ [the so-called Toulouse point]. A subsequent adiabatic RG around $\alpha = 1/2$ might then lead to sensible results. We did not carry out such an analysis in the present work, though.

We pursued our analysis by calculating the RG equations of $\tilde\alpha\equiv\alpha \omega_c^{s-1}$, $\tilde h\equiv 2h/\omega_c$ and $J$ up to third order in $\tilde h$ for the system consisting of two noisy spins coupled \emph{via} an interaction $J$ in the $z$-direction. We assumed the initial $J(0)$ to be of order of $h(0)$. As long as $J(\ell)/h(\ell) \sim 1$ the adiabatic RG group yields the mean-field RG equations for each spin: By setting, e.g., for the first spin $\epsilon = J\sigma_2^z/2 = const.$ we reproduce the RG equations of the single-spin system. 
For small inter-spin coupling $J(\ell)$ the RG equations are given in (\ref{eq:RGJ}) and (\ref{eq:RGJha}). Since the bath cutoff $\omega_c$ decreases during the RG flow, $J(\ell)/\omega_c(\ell)$ does not remain small. In the following large-$J/\omega_c$ analysis we showed that the RG equation of $\tilde h$ remains unchanged whereas the one for $\tilde\alpha$ is heavily modified: For an \emph{Ohmic bath} this leads to the destruction of the Kosterlitz-Thouless transition usually observed in an Ohmic spin-boson system. Instead, the system shows a second order phase transition around the critical effective bath coupling $\alpha^* = 1$. It is not possible within our formalism to exactly determine the dependence of $\alpha^*$ on the initial $\alpha$ since the exact crossover equations between the small $J$ and the large $J$ regimes are not known. Hence, it is also difficult to determine the critical exponents. However, in the limiting case $h(0)/\omega_c(0)\to 0$ the flow of $\alpha(\ell)$ can be neglected. We can then argue that $\alpha_c = 1$ in the two-spin system.  

In the case of a \emph{sub-Ohmic} bath, the adiabatic RG equations predicts a localization of the spins for \emph{all values of $h$} in sharp contrast to the behaviour of the single unbiased sub-Ohmic spin-boson model which has a second-order phase transition with an $h$- and $\omega_c$-dependent $\tilde\alpha_c$. It should be noted that this is a rather unexpected result which is likely to hold for the quantum Ising chain as well.

In~\cite{Cugl05} the authors simulated, amongst others, systems of $N_s= 2$ and $N_s = 4$ noisy spins (for the \emph{Ohmic} case). Their simulation was restricted to values of $\tilde h \ge 0.2$ and the small $\tilde h$-limit could not be attained. They clearly showed that $\alpha_c$ depends on the number of coupled spins as long as $\tilde h \ge 0.2$. However, when $\tilde h$ is further lowered we expect the critical dissipation to reincrease until $\alpha_c(N_s,\tilde h\ll 1)  = 1$ independent of $J$ and of $N_s$ (by extrapolating our results for $N_s=1,2$). The probable scenario is the following: Upon increasing $N_s$, $\alpha_c(N_s,\tilde h)$ decreases for $\tilde h \in [\tilde h_m(N_s),\tilde h_M(N_s)]$ with the lower bound $ \tilde h_m(N_s) \to 0^+$ for $N_s \to \infty$. In the limiting case of the dissipative Ising chain $\alpha_c(\infty,\tilde h\ll 1)$ drops to a $J$-dependent value smaller than one which is for $J\ge h$ equal to zero~\cite{Troyer05}.
 
We are convinced that the direct analytical treatment of the dissipative Ising chain is well within reach by further developing our adiabatic renormalization scheme. To date, the only analytical theory available is the critical dissipative $\phi^4$-theory~\cite{Sachdev04}. 
Note that the RG equations (\ref{eq:RGh22}), (\ref{eq:RGalpha}) and (\ref{eq:RGJ}) give us already an idea of how the phase diagram of the Ising chain, where each spin is coupled to a different \emph{Ohmic bath}, should look like. The isolated chain has a ferromagnetic transition at $h = J$. Then, in the presence of the bath $h$ and $J$ are renormalized. If we assume the analogous relation $h^* \propto J^*$ (where $h^* = h(\infty)$ and $J^* = J(\infty)$ are the fully renormalized couplings) for the ferromagnetic transition \emph{with} site dissipation the initial couplings $\alpha_0$, $h_0$ and $J_0$ have to satisfy 
\begin{equation}
  h^*(\alpha_0,h_0,J_0,\omega_c) \propto J^*(\alpha_0,h_0,J_0,\omega_c)
\end{equation}
at the critical point. By inspecting (\ref{eq:RGh22}) we find that $h^* = 0$ for $\alpha_0 > \alpha_c$, from which we deduce that the bath lowers the critical $J_0$ for fixed $h_0$ until $J_0 = 0$ for $\alpha_0>\alpha_c$. Such a behaviour has indeed been observed in the simulations~\cite{Troyer05}. 

The adiabatic renormalization equations break down as soon as $\omega_c \sim h$ and therefore the low frequency oscillators cannot be dealt with within this approach. However, in the last years several articles have demonstrated that a variational theory of simple trial wave functions can be a promising candidate to find accurate approximations of the spin-boson ground state~\cite{Irish07,Chin06,Chin11}. We think that it is possible to combine such variational calculations with the adiabatic RG, which in turn would yield RG equations valid on the whole range of parameters.

\acknowledgements{
The author thanks in particular L. F. Cugliandolo for most valuable advices and suggestions and a critical reading of the manuscript.}


\begin{appendices}

\section{Degenerate perturbation details}\label{sec:rendetails}

We review in this section the main formulae of degenerate static perturbation theory that we used in this present work. The usual idea is to write a perturbation series in some small parameter $p$ of the full states $|\psi_{n}\rangle = |\tilde n\rangle + |\tilde n^1\rangle + |\tilde n^2\rangle + \cdots$ (with $|\tilde n^k\rangle \sim p^k$, $k>0$, and $|\tilde n\rangle$ the unperturbed state) and of the full energies $\mathcal E_n = E_n + E_n^1 + E_n^2 + \cdots$. $|\tilde n\rangle$ is the true physical superposition of the \emph{a priori} basis vectors $|k\rangle, k\in\mathfrak D$ of $\mathfrak D$. More precisely, we have 
\begin{eqnarray*}
  |\tilde n\rangle &=& \sum_{k\in\mathfrak D} c_{nk} |k\rangle\;,\\
  |\tilde n^1\rangle &=& \sum_{k\in\mathfrak D} c_{nk}^1 |k\rangle
  + \sum_{k \notin\mathfrak D}\langle k|\tilde n^1\rangle |k\rangle\;, 
\end{eqnarray*}
etc. 
We remind the reader that the unperturbed states and energies do not carry a superscript $^0$ for the sake of a clear notation.

The first order Schr\"odinger equation reads
\begin{equation}\label{eq:first}
  \mathcal H_0|\tilde n^1\rangle + \mathcal V|\tilde n\rangle = E_n|\tilde n^1\rangle + E_n^1 |\tilde n\rangle \; .
\end{equation}
The general strategy is to multiply the Schr\"odinger equation [whose first-order approximation is given by (\ref{eq:first})] by $\langle k|$ from the left. For $k\notin\mathfrak D$ we have
\begin{equation}\label{eq:firstorder2}
 \langle k |\tilde n^1\rangle = \frac{\mathcal V_{k\tilde n}}{E_{nk}} = 
\sum_{k_1\in\mathfrak D}c_{n k_1}\frac{\mathcal V_{kk_1}}{E_{nk}} \;.
\end{equation}
We introduced the notation $\mathcal V_{kk'} = \langle k|\mathcal V|k'\rangle$ and $E_{nk} = E_n-E_k$. For $k \in \mathfrak D$ on the other hand we obtain
\begin{equation}
  \sum_{k'\in\mathfrak D}\mathcal V_{kk'}c_{nk'} = E_n^1 c_{nk} \;,
\end{equation}
which shows that $\vec c_n \equiv (c_{nk})$ is the eigenvector with eigenvalue $E_n^1$ of the first-order perturbation operator $V^{(1)} = (\mathcal V_{kk'})$.

By repeating the preceding analysis we obtain the second order result. From the second order Schr\"odinger equation
\begin{equation}
  \mathcal H_0|\tilde n^2\rangle + \mathcal V|\tilde n^1\rangle = E_n|\tilde n^2\rangle + E_n^1 |\tilde n^1\rangle + E_n^2 |\tilde n\rangle\; 
\end{equation}
we have for $k \notin\mathfrak D$
\begin{widetext}
\begin{equation}
  \langle k|\tilde n^2\rangle = \sum_{\substack{k_1\notin \mathfrak D\\k_2\in\mathfrak D}}
\frac{\mathcal V_{kk_1}\mathcal V_{k_1k_2}}{E_{nk}E_{nk_1}}c_{nk_2}
- E_n^1 \sum_{k_2\in\mathfrak D} \frac{\mathcal V_{kk_2}}{E_{nk}^2} c_{nk_2}
= \sum_{\substack{k_1\notin \mathfrak D\\k_2\in\mathfrak D}}
\frac{\mathcal V_{kk_1}\mathcal V_{k_1k_2}}{E_{nk}E_{nk_1}}c_{nk_2}
- \sum_{k_1,k_2\in\mathfrak D} \frac{\mathcal V_{kk_2}\mathcal V_{k_2k_1}}{E_{nk}^2} c_{nk_1}
\end{equation}
\end{widetext}
and for $k \in \mathfrak D$
\begin{equation}
  \sum_{\substack{k_1\notin \mathfrak D \\k_2\in\mathfrak D}}\frac{\mathcal V_{kk_1}\mathcal V_{k_1k_2}}{E_{nk_1}}c_{nk_2} + \sum_{k_2\in\mathfrak D}\mathcal V_{kk_2} c_{nk_2}^1 = E_n^1 c_{nk}^1 + E_n^2 c_{nk} \;,
\end{equation}
which can be rewritten by using the first-order result as 
\begin{equation}
V^{(2)} (\vec c_n + \vec c_n^1) = (E_n^1 + E_n^2)(\vec c_n + \vec c_n^1) \; .
\end{equation}
The second-order perturbation operator is defined as 
\begin{equation}
  V^{(2)}_{kk'} = \sum_{k_1\notin\mathfrak D}\frac{\mathcal V_{kk_1}\mathcal V_{k_1k'}}{E_{nk_1}} + V^{(1)}_{kk'}\;.
\end{equation}
Note that the $n$-th order perturbation operator is the perturbation operator \emph{up to} $n$-th order to use the same terminology as in the main text. By using the results in the main text we know that $c_{nk}^1 = 0$ in our case. This will lead to some simplifications in the following higher-order analysis.

The third-order result can be deduced in the same way as before. 
From the third order Schr\"odinger equation
\begin{equation}
  \mathcal H_0|\tilde n^3\rangle + \mathcal V|\tilde n^2\rangle = E_n|\tilde n^3\rangle + E_n^1 |\tilde n^2\rangle + E_n^2 |\tilde n^1\rangle + E_n^3 |\tilde n\rangle\; 
\end{equation}
we find for $k \notin \mathfrak D$
\begin{widetext}
\begin{eqnarray}\label{eq:n3}
  \langle k|\tilde n^3\rangle &=& \sum_{\substack{k_1,k_2\notin\mathfrak D\\k_3\in\mathfrak D}}\frac{\mathcal V_{kk_1}\mathcal V_{k_1k_2}\mathcal V_{k_2k_3}}{E_{nk}E_{nk_1}E_{nk_2}} c_{nk_3} 
- E_n^1\sum_{\substack{k_1\notin\mathfrak D\\k_2\in\mathfrak D}}\frac{\mathcal V_{kk_1}\mathcal V_{k_1k_2}}{E_{nk}^2 E_{nk_1}}c_{nk_2}
- E_n^1\sum_{\substack{k_1\notin\mathfrak D\\k_2\in\mathfrak D}}\frac{\mathcal V_{kk_1}\mathcal V_{k_1k_2}}{E_{nk} E^2_{nk_1}}c_{nk_2} \nonumber\\
&-& E_n^2 \sum_{k_1\in\mathfrak D}\frac{\mathcal V_{kk_1}}{E_{nk}^2}c_{nk_1}
+ \sum_{k_1\in\mathfrak D}\frac{\mathcal V_{kk_1}}{E_{nk}} c_{nk_1}^2
+ (E_n^1)^2\sum_{k_1\in\mathfrak D}\frac{\mathcal V_{kk_1}}{E_{nk}^3}c_{nk_1}
\end{eqnarray}
\end{widetext}
By multiplying the third-order Schr\"odinger equation by $\langle k|$, where $k \in \mathfrak D$, one finds 
\begin{equation}
  V^{(3)} (\vec c_n + \vec c_n^2) = 
(E_n^1 + E_n^2 + E_n^3)(\vec c_n + \vec c_n^2)\;,
\end{equation} 
where we used $\vec c_n^1 = 0$. The third-order perturbation operator reads
\begin{align}\label{eq:Vthird}
  V^{(3)}_{kk'} &= \sum_{k_1,k_2 \notin\mathfrak D} \frac{\mathcal V_{kk_1}\mathcal V_{k_1k_2}
\mathcal V_{k_2k'}}{E_{nk_1}E_{nk_2}} 
- \sum_{\substack{k_1\notin\mathfrak D\\k_2\in\mathfrak D}} \frac{\mathcal V_{kk_1}\mathcal V_{k_1k_2}
\mathcal V_{k_2k'}}{E^2_{nk_1}} \\
&+ V^{(2)}_{kk'} + V^{(1)}_{kk'}\; .
\end{align}
%

Finally, we wish to deduce the fourth-order perturbation operator. By multiplying the fourth order Schr\"odinger equation
\begin{equation}
  \mathcal H |\tilde n^4\rangle + \mathcal V |\tilde n^3\rangle
= E_n |\tilde n^4\rangle + E_n^1 |\tilde n^3\rangle + E_n^2 |\tilde n^2\rangle
+ E_n^3|\tilde n^1\rangle + E_n^4|\tilde n\rangle
\end{equation}
by $\langle k|,k\in\mathfrak D$, we obtain
\begin{equation}
  \langle k|\mathcal V|\tilde n^3\rangle = E_n^1 c_{nk}^3 + E_n^2 c_{nk}^2 + E_n^3 c_{nk}^1 + E_n^4 c_{nk} \; .
\end{equation}
We now replace $|\tilde n^3\rangle$ by (\ref{eq:n3}) for $k\notin\mathfrak D$ and by $c_{nk}^3$ for $k \in\mathfrak D$ to find, by using the first to third order results,
\begin{equation}
  V^{(4)} (\vec c_n + \vec c_n^2 + \vec c_n^3) = 
(E_n^1 + E_n^2 + E_n^3 + E_n^4)(\vec c_n + \vec c_n^2 + \vec c_n^3)  \; .
\end{equation}
At last, we give the expression for $V^{(4)}$:
\begin{widetext}
\begin{eqnarray}\label{app:V4}
  V^{(4)}_{kk'} &=& \sum_{k_1,k_2,k_4\notin\mathfrak D}
\frac{\mathcal V_{kk_4}\mathcal V_{k_4k_1}\mathcal V_{k_1k_2}\mathcal V_{k_2k'}}{E_{nk_4}E_{nk_1}E_{nk_2}}
- \sum_{\substack{k_2,k_4\notin\mathfrak D\\k_1\in\mathfrak D}}
\frac{\mathcal V_{kk_4}\mathcal V_{k_4k_1}\mathcal V_{k_1k_2}\mathcal V_{k_2k'}}{E^2_{nk_4}E_{nk_2}}
- \sum_{\substack{k_1,k_4\notin\mathfrak D\\k_2\in\mathfrak D}}
\frac{\mathcal V_{kk_4}\mathcal V_{k_4k_1}\mathcal V_{k_1k_2}\mathcal V_{k_2k'}}{E^2_{nk_4}E_{nk_1}} \nonumber\\
&-&  \sum_{\substack{k_1,k_4\notin\mathfrak D\\k_2\in\mathfrak D}}
\frac{\mathcal V_{kk_4}\mathcal V_{k_4k_1}\mathcal V_{k_1k_2}\mathcal V_{k_2k'}}{E_{nk_4}E^2_{nk_1}} 
+   \sum_{\substack{k_4\notin\mathfrak D\\k_1,k_2\in\mathfrak D}}
\frac{\mathcal V_{kk_4}\mathcal V_{k_4k_1}\mathcal V_{k_1k_2}\mathcal V_{k_2k'}}{E_{nk_4}^3}
+ V^{(3)}_{kk'}
+ V^{(2)}_{kk'}
+ V^{(1)}_{kk'}
\;.
\end{eqnarray}
\end{widetext}


\section{Fourth-order double spin-flip matrix element canceling in the small $J$ analysis}

%
We demonstrate here that the fourth order contribution to a double spin-flip matrix element cancels in the case $J/\omega_c \ll 1$. We have
\begin{align}\label{eq:V420}
  \mathcal V_2^{(4)} &= \sum_{k_1,k_2,k_3\notin\mathfrak D}\frac{\langle ++|\mathcal V|k_1\rangle
\langle k_1|\mathcal V|k_2\rangle\langle k_2|\mathcal V|k_3\rangle\langle k_3|\mathcal V|--\rangle}
{-E_{k_1}E_{k_1}E_{k_1}} \nonumber\\ 
&- \sum_{\substack{k_1,k_2\notin\mathfrak D\\k_3\in\mathfrak D}}\frac{\langle ++|\mathcal V|k_1\rangle
\langle k_1|\mathcal V|k_2\rangle\langle k_2|\mathcal V|k_3\rangle\langle k_3|\mathcal V|--\rangle}
{-E_{k_1}^2E_{k_2}} \nonumber
\\
&-\sum_{\substack{k_1,k_2\notin\mathfrak D\\k_3\in\mathfrak D}}\frac{\langle ++|\mathcal V|k_1\rangle
\langle k_1|\mathcal V|k_2\rangle\langle k_2|\mathcal V|k_3\rangle\langle k_3|\mathcal V|--\rangle}
{-E_{k_1}E^2_{k_2}}\nonumber\\
&- \sum_{\substack{k_1,k_3\notin\mathfrak D\\k_2\in\mathfrak D}}\frac{\langle ++|\mathcal V|k_1\rangle
\langle k_1|\mathcal V|k_2\rangle\langle k_2|\mathcal V|k_3\rangle\langle k_3|\mathcal V|--\rangle}
{-E_{k_1}^2E_{k_3}}  \nonumber
\\
&+ \sum_{\substack{k_1\notin\mathfrak D\\k_2,k_3\in\mathfrak D}}\frac{\langle ++|\mathcal V|k_1\rangle
\langle k_1|\mathcal V|k_2\rangle\langle k_2|\mathcal V|k_3\rangle\langle k_3|\mathcal V|--\rangle}
{-E_{k_1}^3} \;.
\end{align}
%
For the sake of a comprehensible notation we introduced 
$|++\rangle \equiv |+_1\rangle|\vec n_{1,<}\rangle|0_1\rangle|+_2\rangle|\vec n_{2,<}\rangle|0_2\rangle$ with an analogue definition for $|--\rangle$.
The expression of $\mathcal V_2^{(4)}$ to lowest order in $\omega_c^{-1}$ reads:
\begin{eqnarray}
  \mathcal V_2^{(4)} &=& \frac{{h'}^4}{4}\sum_{j_1} \frac{-\lambda_{j_1}^2}{\omega_{j_1}^2}
\frac{2}{-\omega_{j_1}^3} +\frac{{h'}^4}{4}\sum_{j_2} \frac{-\lambda_{j_2}^2}{\omega_{j_2}^2}
\frac{2}{-\omega_{j_2}^3} \nonumber\\
&-& 2\frac{{h'}^4}{4}\sum_{j_2} \frac{-\lambda_{j_2}^2}{\omega_{j_2}^2}
\frac{1}{-\omega_{j_2}^3} - 
2\frac{{h'}^4}{4}\sum_{j_1} \frac{\lambda_{j_1}^2}{\omega_{j_1}^2}
\frac{1}{-\omega_{j_1}^3} \nonumber\\
&-& 2\frac{{h'}^4}{4}\sum_{j_1} \frac{-\lambda_{j_1}^2}{\omega_{j_1}^2}
\frac{1}{-\omega_{j_1}^3} - 
2\frac{{h'}^4}{4}\sum_{j_2} \frac{\lambda_{j_2}^2}{\omega_{j_2}^2}
\frac{1}{-\omega_{j_2}^3} \nonumber\\
&+&\frac{{h'}^4}{4}\sum_{j_1} \frac{\lambda_{j_1}^2}{\omega_{j_1}^2}\frac{2}{-\omega_{j_1}^3}
+ \frac{{h'}^4}{4}\sum_{j_2} \frac{\lambda_{j_2}^2}{\omega_{j_2}^2}\frac{2}{-\omega_{j_2}^3}
\; .
\end{eqnarray}
The first line stems from the first summand of the rhs of (\ref{eq:V420}), the second and the third lines from the second summand and the fourth line from the last summand of (\ref{eq:V420}). The fourth summand does not contribute. Note also that only terms which do not leave $\mathcal D$ invariant arise at this order, i.e. only the $h$-term in $\mathcal V$ contributes here. 
The final result thus reads
\begin{equation}\label{eq:V42}
\mathcal V_2^{(4)} =  0 \; ,
\end{equation}
as expected.

\end{appendices}

\bibliographystyle{phjcp}
\bibliography{artbib1.bib}

\end{document}